\documentclass[twocolumn,showpacs,preprintnumbers,amsmath,amssymb,pra]{revtex4}

\usepackage{color}
\usepackage{graphicx}
\usepackage{dcolumn}
\usepackage{bm}

\begin{document}


\title{Entanglement and parametric resonance in driven quantum systems}
\author{V. M. Bastidas$^1$}
\email{victor@itp.physik.tu-berlin.de}
\author{J. H. Reina$^2$}%
 \author{C. Emary$^1$}
  \author{T. Brandes$^1$}

\affiliation{%
$^1$Institut f\"ur Theoretische Physik, Technische Universit\"at Berlin, Hardenbergstr. 36, 10623 Berlin, Germany}%
  \affiliation{%
$^2$Universidad del Valle, Departamento de F\'isica, A. A. 25360, Cali, Colombia
}%

\date{\today}

\begin{abstract}
We study the relationship between entanglement and parametric resonance in a system of two coupled time-dependent oscillators. As a measure of bipartite entanglement,
we calculate the linear entropy for the reduced density operator, from which we study the entanglement dynamics. In particular, we find that the bipartite entanglement increases in time up to a maximal mixing scenario, when the set of auxiliary dynamical parameters are under parametric resonance. Moreover, we obtain a closed relationship between the correlations in the ground state, the localisation of the Wigner function in phase space, and the localisation of the wave function of the total system.

\end{abstract}






\maketitle
%

\section{\label{intro}Introduction}
A quantum system composed of two or more entangled subsystems has the intriguing property that
although the state of the total system can be well defined, it is impossible to identify individual properties for each one of its parts. Understanding the properties displayed by the entanglement of physical systems is one of the fundamental purposes of quantum information theory
\cite{Plenio07,Bennett00}. In this paper we focus on the relationship between entanglement and parametric resonance in a composite quantum system.\\
In classical dynamics, parametric resonance can occur when an appropriate parameter of a system is varied periodically in time \cite{Arnold}. Under parametric resonance, stable points of the undriven system become unstable for specific values of the period of the parameter variation. On the other hand, an unstable point may become a stable one --- a phenomenon known as parametric stabilization.
In the quantum regime, it is well known that a periodically driven system follows a unitary dynamics. 
The dynamics is not closed, however, since the external driving changes the physical parameters of the system, altering its total energy. The usual technique for treating such systems is the Floquet Formalism \cite{Shirley,dittrich}. Perelomov and Popov \cite{Perelomov}, and most recently Weigert \cite{weigert}, have studied quantum manifestations of classical parametric resonance in noncomposite quantum systems in the context of Floquet theory for the Schr\"{o}dinger equation. They found, for example, that for parameters in the stable regions, a discrete spectrum of quasienergies exist. In contrast, in the unstable zones, there is a continuous spectrum.\\
The question that naturally arises is: what are the quantum signatures of classical parametric resonance? There is no classical behavior analogous to entanglement, it is unique to quantum mechanics.
In this paper we show that this intrinsically quantum property shows strong signatures of classical parametric resonance in a system of two driven coupled harmonic oscillators. Furthermore, we find that the entanglement dynamics is related to the dynamical behaviour of the Wigner function and the total system's ground-state wave packet.
This paper is organized as follows: In section \ref{model} we describe the model. In section \ref{LRCM} we introduce the required basic aspects of the Lewis-Riesenfeld canonical
method and the single parametric oscillator. In section \ref{PR} we find the exact solution of the Schr\"{o}dinger equation and discuss the stability properties of the auxiliary dynamical parameters and their relation to the phenomenon of parametric resonance. In section \ref{QDEQ} we present the calculation of the entanglement dynamics and the Wigner function. Finally, discussion of the results and conclusive remarks are presented in sections \ref{D} and \ref{C} respectively.
\section{\label{model}Model}
Time-periodic quadratic Hamiltonians are known to provide examples of parametric resonance.
This type of Hamiltonian has the general form
\begin{equation}
\label{GENHAM}
\hat{H}(t)=\frac{1}{2}\sum_{i=1}^{2}(\mu_{i}(t)\hat{p}_{i}^{2}
+\nu_{i}(t)\hat{x}_{i}^{2})+\gamma(t)\hat{x}_{1}\hat{x}_{2}.
\end{equation}
In this work we are interesting in studying the particular model described by the time-periodic Hamiltonian
\begin{equation}
\label{GENHAMPAR}
\hat{H}(t)=\frac{1}{2}\sum_{i=1}^{2}(\hat{p}_{i}^{2}
+\omega^{2}\hat{x}_{i}^{2})+\gamma(t)\hat{x}_{1}\hat{x}_{2},
\end{equation}
where $\gamma(t)=g+\Delta g \cos \Omega t$, $\omega $ is a time-independent frequency, $g$ is called the static coupling, and $\Delta g$ is a fraction of $g$. \\
Further motivation for studying this type of Hamiltonian comes from the time-dependent Dicke model.
In 1954, Dicke showed, in a now celebrated paper \cite{Dicke}, the coherent spontaneous emission
arising from many atoms emitting collectively.
The single mode Dicke Hamiltonian models the interaction of $N$ atoms with a
single mode bosonic field via dipole interactions within an ideal
cavity. The time-dependent generalization of Dicke model is described by the Hamiltonian
\begin{equation}
\hat{H}(t)= \omega_0(t) J^{z}+ \omega(t)\hat{a}^\dagger \hat{a}+ \frac{\lambda(t)}{\sqrt{N}}
(J^{+}+J^{-})\left(\hat{a}+\hat{a}^\dagger\right) \ .
\label{TDickeHam}
\end{equation}
%
By considering the Holstein-Primakoff transformation of
the angular momentum algebra and taking the limit $j\rightarrow \infty$ in the Hamiltonian Eq. \eqref{TDickeHam}, one obtains an effective Hamiltonian \cite{brandes1,brandes2}, which corresponds to the Hamiltonian of two time-dependent oscillators
linearly coupled through the interaction strength $\lambda(t)$. 
In terms of the field coordinate $x$ and the atoms coordinate $y$, this Hamiltonian becomes
\begin{eqnarray}
\label{eq:hamoa} \hat{H}(t)&=&\frac{\omega(t)}{2
\omega}\left(\hat{p}_{x}^{2}+\omega^{2}\hat{x}^{2}\right)
+\frac{\omega_{0}(t)}{2
\omega_{0}}\left(\hat{p}_{y}^{2}+\omega_{0}^{2}\hat{y}^{2}\right)
+ \nonumber
\\&&  2 \lambda(t)\sqrt{\omega \omega_
{0}}\ \hat{x} \hat{y} -E_{0} \ ,
\end{eqnarray}
where $\omega(t_{0})=\omega$, $\omega_{0}(t_{0})=\omega_{0}$, and $E_{0}(t)=\frac{\omega_{0}(t)}{2}+\frac{\omega(t)}{2}+j\omega_{0}(t)$.
In the next section we describe the Lewis-Riesenfeld canonical method and thus establish the non-perturbative technique used in this work.
%
\section{\label{LRCM}The Lewis--Riesenfeld canonical method and the single parametric oscillator}
In order to study the nonequilibrium dynamics described by time-dependent
Hamiltonians, we resort to the Lewis-Riesenfeld canonical method. This technique deals with the
relation between the eigenstates of an invariant operator and the
solutions of the corresponding Schr\"{o}dinger equation
\cite{lewis}. Using this method, it is possible to find the exact solution
for a single parametric quantum oscillator \cite{lewis, Hidalgo, Kim}.
\subsection{The Lewis--Riesenfeld canonical method}
An explicitly time-dependent invariant operator
$\hat{O}$ should satisfy the condition
\begin{equation}
\label{condition}
\frac{d}{dt}\hat{O}(t)=\frac{\partial}{\partial t}\hat{O}(t)+i[\hat{H}(t),\hat{O}(t)] = 0 \ .
\end{equation}
We write the system's dynamics by means of the standard
Schr\"{o}dinger equation
\begin{equation}
i\frac{\partial}{\partial t}|\psi,t\rangle
=\hat{H}(t)|\psi,t\rangle \ . \label{schro}
\end{equation}
A non-equilibrium system evolves towards a final state which, in
general, differs from the initial one.
Lewis and Riesenfeld observed that any operator $\hat{O}(t)$
satisfying Eq. \eqref{condition} can be used to construct the exact quantum states of  the
Schr\"{o}dinger equation. In general, if $|\psi,t\rangle$ is a
solution, this can be written in terms of the eigenstates of the
operator $\hat{O}(t)$,
\begin{equation}
\label{WFL} |\psi,t\rangle =
\sum_{n}c_{n}\exp(i\ \theta_{n}(t))|\lambda_{n},t\rangle \ ,
\end{equation}
where
$\hat{O}(t)|\lambda_{n},t\rangle=\lambda_{n}|\lambda_{n},t\rangle$,
and  the time dependent phase is $\theta_{n}(t) = \int_{0}^{t}
\langle \lambda_{n},t|i\frac{\partial}{\partial
t}-\hat{H}(t)|\lambda_{n},t\rangle dt$.
\subsection{The single quantum  parametric oscillator}
We consider the Hamiltonian
\begin{equation}
\label{eq:hamoda}
\hat{H}(t)=\frac{1}{2}\hat{p}^{2}+\frac{\omega^{2}(t)}{2}\hat{q}^{2} \ ,
\end{equation}
where $\omega(t)$ is an arbitrary (not necessarily real), piecewise-continuous function of time \cite{lewis}.
For this system, one can derive a set of time dependent invariants which are homogeneous, quadratic expressions in the dynamical variables $p$ and $q$. For this purpose, we introduce time dependent ladder operators $\hat{a}(t)$ and $\hat{a}^{\dagger}(t)$, that satisfy Eq. \eqref{condition}. For example, the annihilation operator is given by
\begin{equation}
\label{aux0}
\hat{a}(t)=-\dot{B}(t)\hat{q}+B(t)\hat{p} \ ,
\end{equation} 
where the auxiliary dynamical parameter $B(t)$ satisfy the classical equation of motion
\begin{equation}
\label{aux1}
 \ddot{B}(t)+\omega^{2}(t) B(t)=0 \ ,
\end{equation}
and the Wronskian condition
\begin{equation}
\label{aux2}
\dot{B}(t)B^{\ast}(t)-B(t)\dot{B}^{\ast}(t)=i \ .
\end{equation}
The last condition ensure that the operators are ladder operators at all times \cite{Hidalgo,Kim}. In the context of Lewis-Riesenfeld's canonical method, we construct the Hermitian invariant operator $\hat{n}(t)=\hat{a}^{\dagger}(t)\hat{a}(t)$ and the Fock space consisting of time-dependent number states
\begin{equation}
\hat{n}(t)|n,t \rangle =n|n,t \rangle \ ,
\end{equation}
where
\begin{equation}
\label{parfock}
|n,t\rangle=\frac{(\hat{a}^{\dagger}(t))^{n}}{\sqrt{n!}}
|0,t \rangle \ ,
\end{equation}
and the vacuum state $|0,t\rangle$ is anihilated by  $\hat{a}(t)$.\\
The dynamical behaviour of the system is fully determined through the solution 
of the corresponding classical problem Eq. \eqref{aux1}.
\section{\label{PR}The Lewis--Riesenfeld canonical method for coupled oscillators and classical parametric resonance}
In this section we find the solution of the Schr\"{o}dinger equation for
the Hamiltonian Eq. \eqref{GENHAMPAR} and study the stability properties of the auxiliary dynamical
parameters, which are related to
the phenomenon of parametric
resonance in classical mechanics \cite{Arnold}.
\subsection{Exact solution of the Schr\"odinger equation}
In the case of the Schr\"{o}dinger equation for
the Hamiltonian Eq. \eqref{GENHAMPAR}, there is a transformation that allows us to map this Hamiltonian into a
Hamiltonian of two uncoupled parametrically driven quantum
harmonic oscillators.
We define the time-independent unitary transformation $\hat{U}$ acting
over the system's state $|\Psi,t\rangle$, to obtain
\begin{equation}
\label{eq:tus} |\Psi,t\rangle_{R} =\hat{U}|\Psi,t\rangle \ .
\end{equation}
We choose the unitary operator $\hat{U}$ such  that the system's Hamiltonian $\hat{H}(t)$ is taken to a new operator which can be  associated to that of two uncoupled time dependent quantum harmonic
oscillators, $\hat{H}_{R}(t)=\hat{U}^{\dagger}\hat{H}(t)\hat{U}$, such that
\begin{equation}
\label{eq:hamoda}
\hat{H}_{R}(t)=\frac{1}{2}\left(\hat{p}_{-}^{2}+(\epsilon_{-}(t))^{2}\hat{q}_{-}^{2}+\hat{p}_{+}^{2}+(\epsilon_{+}(t))^{2}\hat{q}_{+}^{2}\right) \ ,
\end{equation}
where the energies of the two independent oscillator modes
$\epsilon_{\mp}(t)$ are defined in terms of the other control
parameters as $\left(\epsilon_{\mp}(t)\right)^{2}=\omega^{2}\mp
2\omega|\gamma(t)|$. The solution of the Schr\"odinger equation for the
Hamiltonian of Eq. \eqref{eq:hamoda} is obtained as a linear
combination of product of quantum states
\begin{equation}
\label{eq:solbs} |\Psi_{n_{-},n_{+}},t\rangle_{R}=\exp
i(\theta_{n_{-}}(t)+\theta_{n_{+}}(t))|n_{-},t\rangle_{R}
|n_{+},t\rangle_{R} \ ,
\end{equation}
where $|n_{\mp},t\rangle_{R}$ are the quantum states corresponding
to the oscillator mode with excitation energy $\epsilon_{\mp}(t)$.
The fundamental excitations of the system are given by the
energies $\epsilon_{\mp}(t)$. The general solution of the Schr\"odinger equation is
given by
\begin{equation}
\label{eq:sols} |\Psi,t\rangle_{R} =
\sum_{n_{-},n_{+}}c_{n_{-},n_{+}}|\Psi_{n_{-},n_{+}},t\rangle_{R} \ .
\end{equation}
In order to find the wave functions corresponding to the quantum
states $|n_{\mp},t\rangle_{R}$, we now go back to the Lewis-Riesenfeld's
canonical method. The first step in the construction of the
solution is to find the ground state of the system, which must
satisfy the condition
\begin{equation}
\label{opereq}
 \hat{a}_{\mp}|0,0,t\rangle_{R} = 0 \ ,
\end{equation}
where $\hat{a}_{\mp}(t)=-\dot{B}_{\mp}(t)\hat{q}_{\mp}+B_{\mp}(t)\hat{p}_{\mp}$
are annihilation operators corresponding to the oscilator mode
with energy $\epsilon_{\mp}(t)$ that satisfy Eq. \eqref{condition}, and
the auxiliary dynamical parameters $B_{\mp}(t)$ satisfy the
Mathieu equation \cite{abram}
\begin{equation}
\label{eq:eqmath}
 \ddot{B}_{\mp}(t)+(\epsilon_{\mp}(t))^{2} B_{\mp}(t)=0 \ ,
\end{equation}
subject to the Wronskian condition
\begin{equation}
\label{eq:eqwr}
\dot{B}_{\mp}(t)B_{\mp}^{\ast}(t)-B_{\mp}(t)\dot{B}_{\mp}^{\ast}(t)=i\ .
\end{equation}
We next solve the differential equations subjected
to an initial  thermal equilibrium condition
\cite{brandes1,brandes2} satisfying the Wronskian constraint; this
requires that
\begin{eqnarray}
B_{\pm}(t_{0})&=&\frac{i}{\sqrt{2(\epsilon_{\pm}(t_{0}))^{\ast}}}
\ ,\nonumber
\\
\dot{B}_{\pm}(t_{0})&=&-\sqrt{\frac{\epsilon_{\pm}(t_{0})}{2}}\ .
\label{e11}
\end{eqnarray}
In the coordinate representation, the operator
Eq. \eqref{opereq} is equivalent to the differential equation
\begin{equation}
\label{eq:eqgs}
\left(-\dot{B}_{\mp}(t)q_{\mp}+\frac{1}{i}B_{\mp}(t)\frac{\partial}{\partial
q_{\mp}}\right)\Phi_{0}^{\mp}(q_{\mp},t) = 0\ .
\end{equation}
The solution to this equation is the wave function
\begin{equation}
\label{eq:solgs} \Phi_{0}^{\mp}(q_{\mp},t)= \left(\frac{1}{2\pi
|B_{\mp}(t)|^{2}}\right)^{1/4}\exp\left(i\frac{\dot{B}_{\mp}(t)}{2
B_{\mp}(t)} q_{\mp}^{2}\right)\ ,
\end{equation}
thus obtaining the normalized ground state wave function
$\Phi^{R}_{0,0}(q_{-},q_{+},t)=\Phi_{0}^{-}(q_{-},t)\Phi_{0}^{+}(q_{+},t)$.
The construction of the Fock space consisting of time-dependent
number states is direct, as any number state is obtained by applying
the creation operators to the vacuum state
\begin{equation}
|n_{-},n_{+},t\rangle_{R}=\frac{(\hat{a}_{-}^{\dagger}(t))^{n_{-}}}{\sqrt{(n_{-})!}}
\frac{(\hat{a}_{+}^{\dagger}(t))^{n_{+}}}{\sqrt{(n_{+})!}}|0,0,t\rangle_{R}\ .
\end{equation}
In the coordinate representation, the number state is given by
$\Phi^{R}_{n_{-},n_{+}}(q_{-},q_{+},t)=\Phi_{n_{-}}^{-}(q_{-},t)\Phi_{n_{+}}^{+}(q_{+},t)$,
where the quantum states $\Phi_{n_{\mp}}^{\mp}$ can be written in
terms of the Hermite polynomials $H_{n}$ \cite{Kim} as
\begin{eqnarray}
\label{eq:sols} \Phi_{n_{\mp}}^{\mp}(q_{\mp},t)&=&
\left(\frac{2^{-2(n_{\mp})}}{2\pi
|B_{\mp}(t)|^{2}(n_{\mp})!^{2}}\right)^{1/4}
\left(\frac{B^{\ast}_{\mp}(t)}{B_{\mp}(t)}\right)^{n_{\mp}}\times
\nonumber
\\&&
H_{n}(Q_{\mp}(t))
     \exp\left(i\frac{\dot{B}_{\mp}(t)}{2 B_{\mp}(t)}
q_{\mp}^{2}\right)\ ,
\end{eqnarray}
where we have introduced the parameter $Q_{\mp}(t)=(2
|B_{\mp}(t)|^{2})^{-1/2}q_{\mp}$. In analogy to the harmonic
oscillator case, we define a characteristic system length scale
in terms of the auxiliary dynamical parameters as
\begin{equation}
\label{length}
l^{\mp}(t)= \sqrt{2} |B_{\mp }(t)| \ .
\end{equation}
\begin{figure*}
\includegraphics[scale=0.27]{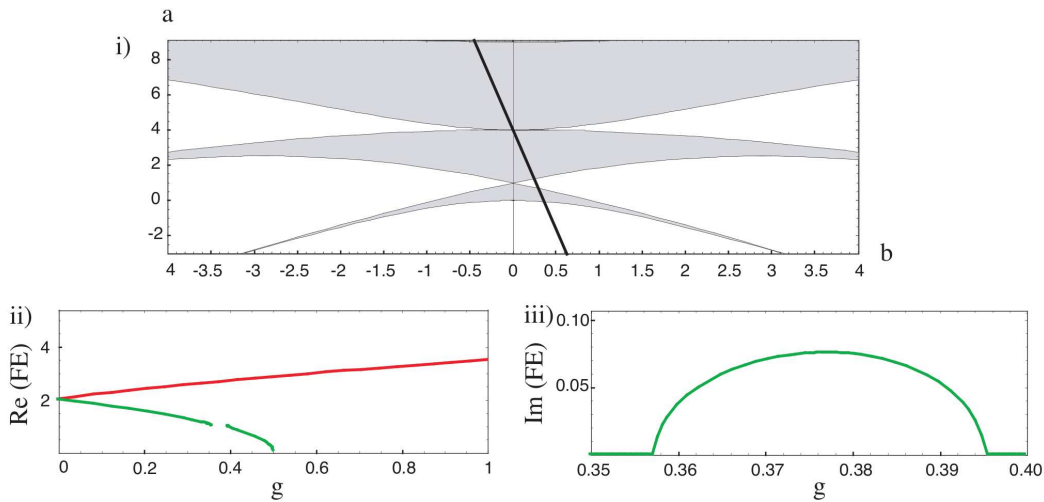}%
\caption{\label{figu1} \small{Stability of the Mathieu's equation:
i) shaded areas correspond to a real Floquet exponent, and  the
white areas to a complex FE; ii) the upper curve shows the real
part of the FE for the solution $B_{+}(t)$ whereas the lower curve
gives the real part of the FE for the solution $B_{-}(t)$; iii)
imaginary part of  the Floquet exponent for the solution
$B_{-}(t)$ in the gap zone of the lower curve in ii).}}
\end{figure*}
The solution of the Schr\"odinger equation for the original Hamiltonian
is obtained by applying the inverse of the unitary transformation
$\hat{U}$ to the quantum state Eq. \eqref{eq:sols}. Finally,
the state of the system $\langle x_{1},x_{2}|\Psi_{n_{-},n_{+}},t\rangle$
can be written as a function of the coordinates $x_{1}$ and $x_{2}$ as follows
\begin{widetext}
\begin{equation}
\label{eq:solcs}
\Psi_{n_{-},n_{+}}(x_{1},x_{2},t)=
\exp(i\ \theta_{n_{-}}(t))\exp(i\ \theta_{n_{+}}(t))
\Phi_{n_{-}}^{-}\left(\frac{x_{1}}{\sqrt{2}}-\frac{x_{2}}{\sqrt{2}},t\right)
 \Phi_{n_{+}}^{+}\left(\frac{x_{1}}{\sqrt{2}}+\frac{x_{2}}{\sqrt{2}},t\right)\ .
\end{equation}
\end{widetext}
\begin{figure*}
\includegraphics[scale=0.37]{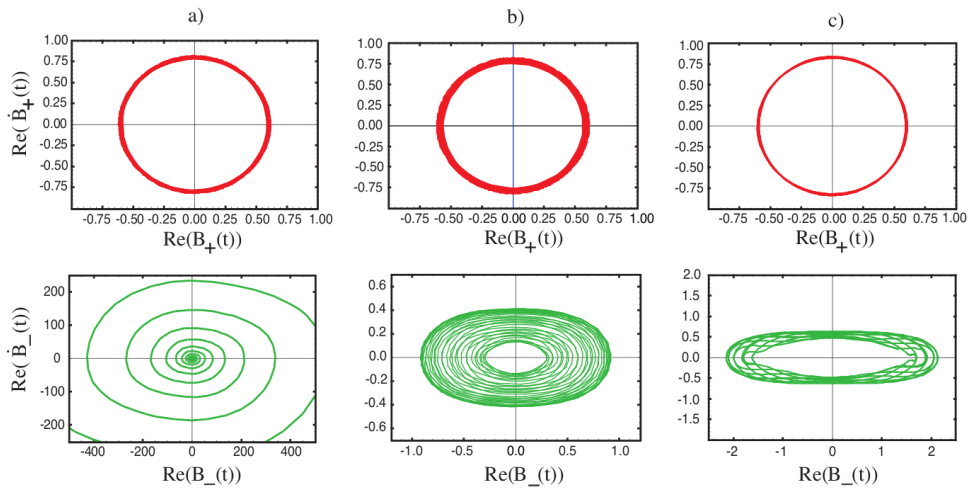}%
\caption{\label{figu2}\small{ Phase space trajectories of the
auxiliary dynamical parameters $B^{\pm}(t)$ for $\Delta g =0.1 g$,
and a value of the static coupling in a) the unstable region,
$g=0.38$, b) the stable region, $g=0.4$, and  c)  the stable
region, $g=0.46$.}}
\end{figure*}
\subsection{Mathieu's equation and parametric resonance}
Mathieu's equation \cite{abram} is given by
\begin{equation}
\label{canmathieu}
\ddot{f}(t)+(a-2b\cos (2t))f(t)=0 \ ,
\end{equation}
the behaviour of its solution depends strongly on the parameters $a$ and $b$, the driving can stabilize
or destabilize the undriven oscillation, this means that the solution
can be bounded or increasing with time respectively. In general, parametric 
resonance can occur if the parameters of a classical dynamical system vary periodically with
time. Stable fixed points of the flow in phase space become
unstable for specific values of certain parameters \cite{Arnold}.
%
Interestingly, the set of differential
equations \eqref{eq:eqmath} has the general form of Mathieu's
equation \cite{abram}
\begin{equation}
\label{mathieu} \ddot{B}_{\mp}(t)+[(\omega^{2} \mp 2 \omega
g)\mp(2\omega\Delta g)\cos \Omega t]B_{\mp}(t)=0 \ .
\end{equation}
If we compare this with the canonical form of the Mathieu's equation of
Eq. \eqref{canmathieu}, it is clear that the parameters $a$ and $b$ are 
functions of the static coupling $g$,
the frequency $\omega$, and the external driving frequency
$\Omega$:
\begin{eqnarray}
 a^{\mp}(g,\omega,\Omega)&=&4(\omega^{2}\mp2 \omega g)/\Omega^{2} \,,\\
b^{\mp}(g,\omega,\Omega)&=&\pm 4 \omega\Delta g/\Omega^{2} \ .
\end{eqnarray}
In order to explore the relevant characteristics of the solutions
of Eq. \eqref{mathieu}, we apply the Floquet theorem for second
order differential equations with time periodic coefficients
\cite{floquetorig}. For this, the solutions of  Eq.
\eqref{mathieu} have the general form
\begin{equation}
B_{\mp}(t)= \exp (i F_{\mp} t)\phi_{\mp}(t) \ ,
\end{equation}
where $\phi_{\mp}(t+T)=\phi_{\mp}(t)$ and
$F_{\mp}(a^{\mp}(g,\omega,\Omega),b^{\mp}(g,\omega,\Omega))$ is
the Floquet exponent (FE), which depends on the shape of the
driving.
The stability zones of the Mathieu's equation are well known: for
driving functions for which $F_{\mp}$ is complex, the solution
becomes unstable, and in the stable regime, $F_{\mp}$ is a real
number \cite{abram,kohler}. In Fig.~\ref{figu1} $(i)$ we show the
stability zones of the Mathieu's equation in the $a-b$ plane:
within the shaded areas the FE is real, whereas within the white
areas this becomes complex. Once we have established the
parameters $\Delta g$, $\omega$, and $\Omega$, the map
$\zeta^{\mp}(g)=(a^{\mp}(g,\omega,\Omega),b^{\mp}(g,\omega,\Omega))$
describes a straight line in the $a-b$ plane, as can be observed
in Fig.~\ref{figu1} $(i)$. The real part $\Re
(F_{\mp}(\zeta^{\mp}(g)))$ of the FE for the auxiliary dynamical parameters
$B^{\mp}(t)$ as a function of the static coupling, exhibits {\it
band-like} regions that correspond to a complex FE of the
solution, which means that this solution is unstable within the
``\textit{gaps}", where the Floquet exponent is imaginary. In
order to study such features, we first investigate the particular
case $\Delta g =0.1 g$, $\omega=1$, and $\Omega = 1$. Fig.~\ref{figu1} $(ii)$ depict the real part of the FE for
the solution $B^{\mp}(t)$ and Fig.~\ref{figu1} $(iii)$ the
imaginary part of FE for the solution $B^{-}(t)$. In these
figures, we find that for this particular case only exist one {\it
band-like} region for the solution $B^{-}(t)$.
In Fig.~\ref{figu2}, we plot the phase space representation of the
trajectories of the auxiliary dynamical parameters $B_{\mp}(t)$.
Figure~\ref{figu2}  $(a)$ shows that when the static coupling
belongs to the instability zone ($g=0.38$) of $B_{-}(t)$, the
solution $B_{+}(t)$ is bounded, while the solution $B_{-}$ is
unbounded in the phase space. In contrast, in
Figs.~\ref{figu2}$(b)$ and ~\ref{figu2} $(c)$, it is clear that
for values of $g$ that belong to a mutual stability zone ($g=0.4$,
and $g=0.46$) the solutions are bounded in the phase space. \\
We can find an exact solution to the Schr\"{o}dinger equation for the general problem described by the Hamiltonian Eq. \eqref{GENHAM} through the Lewis and Riesenfeld canonical method \cite{lewis, Hidalgo, Kim}, the general solution is given in Appendix A.
\section{\label{QDEQ}Quantum dynamics and entanglement quantifiers}
In this section we establish a connection between the non-equilibrium
entanglement generated by the time-dependent coupling  and the behaviour of the
auxiliary dynamical parameters. Furthermore, in order to study the
correlations in the ground state and their relation with the localisation of the total system's wave function, we derive analytical expressions for the reduced density operator of one oscillator and the corresponding Wigner function.
\begin{figure*}
\includegraphics[scale=0.31]{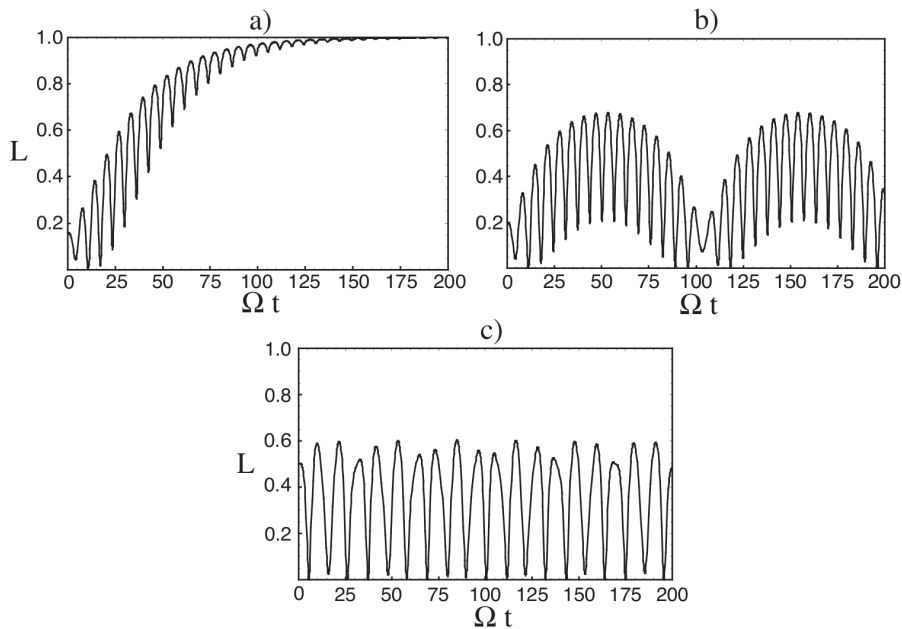}%
\caption{\label{figu3} \small{Linear entropy as a function of time
for $\Delta g =0.1 g$; a) $g=0.38$, b) $g=0.40$, and c) $g=0.46$.
For the case of periodic atom-field coupling.}}
\end{figure*}
\subsection{The reduced density operator}
When studying a composed quantum system whose dynamics is unitary,
it is often more interesting to study a subsystem of the whole
system, which is described by its reduced system dynamics. In
contrast to the dynamics of the whole system, the dynamics of the
subsystem is not unitary. It is possible, however, to make such a
description of the subsystem through the reduced density operator.
We describe the dynamics of the total system in terms of the
density operator, where the pure state of the universe is
represented by the operator $\hat{\rho}_{G}(t)= |\Psi_{0_{-},0_{+}},t \rangle
\langle \Psi_{0_{-},0_{+}},t|$. In the coordinate representation,
the density matrix takes the form
\begin{equation}
\label{eq:opdentot} \rho_{G}(x'_{1},x'_{2};x_{1},x_{2},t)=\Psi
^{\ast}_{0_{-},0_{+}}(x_{1},x_{2},t)\Psi_{0_{-},0_{+}}(x'_{1},x'_{2},t)\ .
\end{equation}
At this stage, we develop a bipartite decomposition of the
universe, which lets us study the reduced dynamics of one oscillator
through the reduced density matrix (RDM). In so doing, we
calculate the partial trace over the physical field coordinate
$x_{1}$,
\begin{equation}
\label{eq:opdenred}
\rho^{(R)}_{G}(x'_{2},x_{2},t)=\int_{-\infty}^{+\infty} \Psi
^{\ast}_{0_{-},0_{+}}(x_{1},x_{2},t)\Psi_{0_{-},0_{+}}(x_{1},x'_{2},t) \ dx_{1}.
\end{equation}
The ground state of the universe
$\Psi_{0_{-},0_{+}}(x_{1},x_{2},t)$ given in Eq. \eqref{eq:solcs} is
Gaussian, which facilitates the integration of Eq. \eqref
{eq:opdenred}. After some algebraic calculations, we obtain the
following result for the reduced density matrix
%
%
%
\begin{eqnarray}
\label{reddenopgen} \rho^{(R)}_{G}(x'_{2},x_{2},t)&=& \Lambda \exp(-\Re
(\alpha)((x'_{2})^{2}+x^{2}_{2})+\beta x_{2} x'_{2}) \times\nonumber
\\&&\,\exp(i\ \Im
(\alpha)((x'_{2})^{2}-x^{2}_{2}))),
\end{eqnarray}
where the parameters $\alpha$, $\beta$, and $\Lambda$ are defined
as
\begin{eqnarray}
\label{paratoms} \nonumber \alpha &=&
\frac{(\xi^{-})^{\ast}s^{2}+(\xi^{+})^{\ast}c^{2}}{2}-\frac{c^{2}s^{2}[(\xi^{-}-\xi^{+})^{\ast}]^{2}}{4(\Re
 (\xi^{-})c^{2}+\Re  (\xi^{+})s^{2})}  , \nonumber \\
\beta&=&\frac{c^{2}s^{2}(\xi^{-}-\xi^{+})^{\ast}(\xi^{-}-\xi^{+})}{2(\Re
 (\xi^{-})c^{2}+\Re  (\xi^{+})s^{2})} ,\nonumber \\
 \Lambda&=&\left(\frac{\Re  (\xi^{-})\Re 
(\xi^{+})}{\pi(\Re  (\xi^{-})c^{2}+\Re 
(\xi^{+})s^{2})}\right)^{1/2} \ ,
\end{eqnarray}
where $c=s=1/\sqrt{2}$  and  $\Re  (\xi^{\mp})$ is the real part of
the function $ \xi ^{\mp}(t)=-i\frac{\dot{B}_{\mp}(t)}{B_{\mp}(t)}$.  In the particular case $\gamma(t)=g$, this function is
independent of time and becomes
\begin{equation}
\label{enegind}
\xi ^{\mp}(t) =
\epsilon_{\mp}=\sqrt{(\omega^{2} \mp 2\omega g)} \ ,
\end{equation}
and corresponds to the result previously reported in the literature for the
atom's reduced density matrix for the single mode Dicke model in
thermal equilibrium \cite{brandes1,brandes2}.
The reduced density operator Eq. \eqref{reddenopgen} has the same
structure as the density operator of an ensemble of
time-dependent oscillators in the coordinate representation
\cite{Kim}.
%
%
\subsection{The linear entropy}
A density operator $\hat{\rho}$  describes a \textit{pure state}
if and only if it satisfies $\hat{\rho}^{2}=\hat{\rho}$, i.e., the
density operator is idempotent. We consider a pure state of a composed system (the universe), say  system $AB$,
represented by the density operator $\hat{\rho}_{AB}$. The linear
entropy for the reduced density operator
$\hat{\rho}_{A}=tr_{B}(\hat{\rho}_{AB})$ of subsystem A is
defined  by  $L_{A}=1-tr_{A}(\hat{\rho}^{2}_{A})$ \cite{Plenio07}.
This gives a measure of purity of the reduced density operator
$\hat{\rho}_{A}$  (one part of the total
system, in a bipartite decomposition of the universe).
If the pure state of the universe is separable, the reduced
density operator of one part of the system represents a pure state
and, as a result, its linear entropy must be zero. Similarly, if the
pure state of the universe is a maximally entangled state
\cite{Nielsen}, the linear entropy equals $1/2$. In the context of our model, it is possible to
use the linear entropy as a measure of bipartite entanglement
for the oscillators system.
The linear entropy for the oscillator's time
dependent reduced density operator, Eq. \eqref{reddenopgen},
is given by $L(t)=1-tr\left[\left(\hat{\rho}^{(R)}_{G}(t)\right)^{2}\right].$ By using the density matrix representation of Eq. \eqref{reddenopgen}, we explicitly calculate  the linear entropy as
\begin{equation}
L(t)=1-\int_{-\infty}^{+\infty}\int_{-\infty}^{+\infty}\rho^{(R)}_{G}(x'_{2},x_{2},t)\rho^{(R)}_{G}(x_{2},x'_{2},t)
\ dx_{2} \ dx'_{2}, \nonumber
\end{equation}
which gives, after some algebra, the result
\begin{equation}
L(t,g)=1-\frac{\pi \Lambda^{2}}{\sqrt{4(\Re 
(\alpha))^{2}-\beta^{2}}} \ . \label{entr1}
\end{equation}
In Fig.~\ref{figu3} we plot  this quantity as a function of time:  it is clear that in the case of a time periodic
atom-field coupling there exist a direct correspondence  between
the stability of the auxiliary dynamical parameters and the
entanglement dynamics, which is reflected in the
graph~\ref{figu3}(a) for a value of the static coupling in the
unstable zone, it is observed that the linear entropy oscillates,
before  reaching the stationary state with \textit{diverging}
linear entropy $L=1$. We consider the linear entropy in
Figs.~\ref{figu3} $(b)$, and~\ref{figu3} $(c)$ for values of the
static coupling in the stable zone.
Our choice of the initial
conditions for Eq.~\eqref{eq:eqmath} implies that the system
starts in thermal equilibrium, but this is described by a
non-separable quantum state. The dynamics exhibits a behaviour
whereby the system experiences a disentanglement process, before
reaching a maximum value of entanglement, and then successive
collapses and revivals. The frequency of the oscillations in the
``stable zones" depends strongly on the value of the static
coupling $g$.
\subsection{The Wigner function}
The Wigner function is a
particular representation of the density operator for a pure or a
mixed state of a quantum system. In principle, the density
operator is more fundamental than its Wigner representation,
however, the Wigner function is often a useful tool in decoherence
studies for investigating possible correlations between position and
momentum \cite{dittrich,zurek,Giulini}.
\begin{figure*}
\includegraphics[scale=0.39]{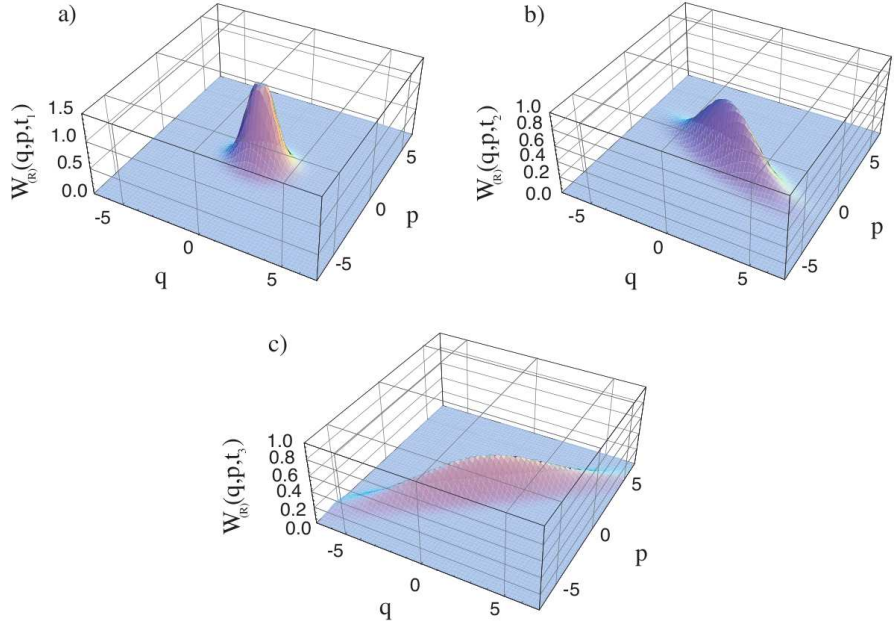}%
\caption{\label{figu4}\small{Time evolution of the Wigner function
in $p-q$ space for the ``unstable zone" $g=0.38$, and $\Delta g
=0.1 g$; a) $\Omega t_{1}=0$, b) $\Omega t_{2}=32$, and c) $\Omega
t_{3}=50$.}}
\end{figure*}
\begin{figure*}
\includegraphics[scale=0.38]{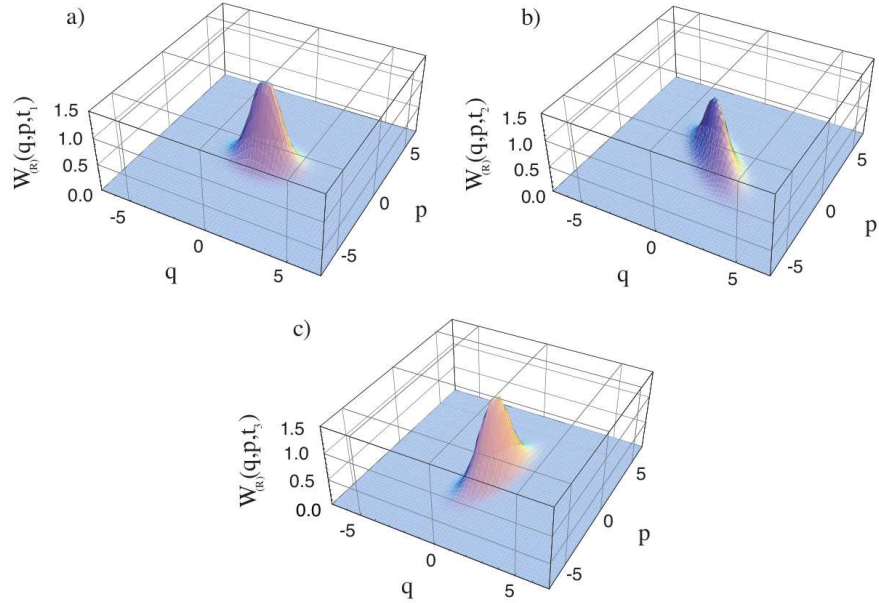}%
\caption{\label{figu5}\small{Time evolution of the Wigner function
in $p-q$-space for the ``stable zone" $g=0.4$, and $\Delta g
=0.1$; a) $\Omega t_{1}=0$, b) $\Omega t_{2}=32$, and c) $\Omega
t_{3}=50$.}}
\end{figure*}
The Wigner representation of a given
density operator $\hat{\rho}$ is given by
\begin{equation}
W(q,p)=\frac{1}{2 \pi}\int_{-\infty}^{\infty} \exp
\left({i p s}\right) \langle
q-s/2|\hat{\rho}|q+s/2\rangle \ ds ,
\end{equation}
where we have considered $\hbar=1$. The Wigner function can take on negative as positive values. This means that it cannot be strictly interpreted  as a phase space
probability density. In the special case when the density operator
represents a pure Gaussian state, the Wigner function is Gaussian and hence is positive definite.
The reduced Wigner function represents the partial trace of a
density operator over a subsystem and so contains all
information about a given subsystem. Here we calculate the  Wigner representation of the oscillator's reduced density operator  given by Eq. \eqref{reddenopgen}. Such Wigner function reads
\begin{equation}
W_{(R)}(q,p,t)=\frac{1}{2 \pi}\int_{-\infty}^{\infty}
e^{ips} \rho^{(R)}_{G}(q-s/2,q+s/2,t) \ ds \ .
\end{equation}
By performing the Gaussian integral we obtain
\begin{widetext}
\begin{eqnarray}
W_{(R)}(q,p,t)=\Lambda \sqrt{\frac{1}{\pi(2\Re  (\alpha)+\beta)}}\exp \left[\frac{4 \Im  (\alpha)
q p}{2\Re (\alpha)+\beta}\right] \exp \left[-\frac{q^{2}(4|\alpha|^{2}-\beta^{2})+p^{2}}{2\Re  (\alpha)+\beta}\right] \ ,
\end{eqnarray}
\end{widetext}
where the parameters $\alpha$, $\beta$, and $\Lambda$ are defined as
in Eq. \eqref{paratoms}.
If we specialize to the case $\gamma(t)=g$, the Wigner
function becomes time-independent,
\begin{equation}
W_{(R)}(q,p)=\left(\frac{2
\sqrt{\epsilon_{-}\epsilon_{+}}}{\pi(\epsilon_{-}+\epsilon_{+})}\right)\exp
\left[-\frac{2}{\epsilon_{-}+\epsilon_{+}}(\epsilon_{-}\epsilon_{+}q^{2}+p^{2})
\right] \ ,
\end{equation}
where $ \epsilon_{\mp}$ are defined in Eq. \eqref{enegind}.
This function exhibits some interesting features: for values of the coupling near to the critical coupling
$ g_{c}=\frac{\omega}{2}$, the Wigner function becomes stretched along the position axis with a consequent
contraction along the momentum axis in the phase space; this results in the Wigner function being delocalized
in phase space.  This behaviour is due to the existing entanglement, because the oscillator's state
becomes a mixed state. Such a behaviour is corroborated by a divergence of the von Neumann entropy 
near  the critical coupling, signaling  a ``maximal mixing" scenario, as reported in Ref. \cite{brandes2} in the context of the time-independent Dicke model.
Figures~\ref{figu4} and
\ref{figu5} show the evolution of the Wigner function in the case
of a periodic atom-field coupling for values of the static
coupling that belong to unstable and stable zones, respectively. In 
our dynamical coupling scenario ($\Delta g\neq 0$), there is a
direct relationship between the atom-field entanglement and the
localisation of the Wigner function in phase space: the Wigner
function evolves in a delocalised region of the phase space for
values of the static coupling in the unstable zone;  as shown in
Fig. \ref{figu4}, the function evolves in such a way that it is
stretched along a dynamical rotating axis with a consequent
contraction in the perpendicular axis, in contrast to the
time-independent behaviour. The behaviour exhibited by the Wigner function
for a value of the static coupling in the stable zone is quite
different: the Wigner function also becomes stretched along a
dynamical axis but it is now localised as the system evolves, as
shown in Fig. \ref{figu5}.
\section{\label{D}Discussion}
From our analytical calculations, we have found that in contrast to the time independent case, the linear
entropy exhibits a time dependence that is not determined by the
global phase of universe's ground state wave function,
Eq.~\eqref{eq:solcs}, but by the auxiliary dynamical parameters
$B_{\mp}(t)$. 
Numerical results suggest that when the set of auxiliary dynamical parameters are under parametric resonance, the trace of $\left(\hat{\rho}^{(R)}_{G}(t)\right)^{2}$  decays exponentially with time with periodic modulations. A careful exploration of the instability
zone, shows that the imaginary part of the Floquet exponent grows
from zero within a certain interval and then falls to zero as
shown in Fig.~ \ref{figu1}$(iii)$. This fact involves a
transition between \textit{oscillatory} and \textit{diverging}
linear entropy and is related with the upper quantum Lyapunov exponent \cite{Jauslin, Fonseca}.
The characteristic length of Eq. \eqref{length} is bounded in the stable zones and unbounded
in the unstable zones. The product of the two
characteristic length scales in the system is proportional to
the Gaussian normalisation factor of the ground state, a result
that tells us about  the relative volume in coordinate space that
the wave packet occupies.
The numerical simulations of the ground state
probability density show that for values of the static coupling
that belong to the common stability zones, the probability is
localized in the ${x}_{1}-{x}_{2}$ space and exhibits an oscillatory
behaviour. In contrast, for values of the static coupling that
belong to the unstable zones, the probability density has also an
oscillatory behaviour but with a different ingredient: when the system evolves, the
density is systematically stretched in a fixed direction, with a
consequent dilation in the perpendicular direction, this behavior results in a massive delocalisation of the wave function of the total system. This behaviour is very suggestive,
because the Wigner function for one oscillator in our case exhibits a similar pattern,
but with a corresponding change in the stretching direction in phase space as time evolves, the Wigner
function of one oscillator ensemble results in a delocalisation in phase space,  a fact that we
associate with the ``maximal mixing" scenario. \\
Recently, Blume-Kohout and Zurek \cite{Blume} analyzed a model in which they considered a harmonic oscillator coupled to an unstable environment, which consisted of an inverted harmonic oscillator. They demonstrated that this unstable environment could produce decoherence in the system of interest more effectively than a bath of harmonic oscillators. They used the von Neumann entropy as a measure of bipartite entanglement, and surprisingly, entropy increases linearly over time with periodic modulations. In contrast to this result, when the environment is a stable oscillator, the von Neumann entropy oscillates, but does not increase over time.
We can think of our system from another perspective: assuming that the universe is composed of two coupled oscillators, we consider that one oscillator is the system of interest, and the other oscillator is the environment. In contrast to the model \cite{Blume}, in our case, both systems are stable, however, control over the dynamics is established through the time-dependent coupling between the oscillators. Numerical results suggest that the von Neumann entropy for the system's reduced density operator exhibits the same behavior described above.
Comparing our results with those found by Blume-Kohout and Zurek, we find that the effect of external control over the system is to produce instability in the universe, which is related to the phenomenon of parametric resonance. This instability is reflected in a loss of coherence in the system of interest.
\section{\label{C}Conclusions}
We have obtained exact results for the oscillator's reduced density
operator and the linear entropy. The results allow us to
study the entanglement dynamics through the calculation of the auxiliary dynamical
parameters. Our results for the Wigner function are to be  contrasted with the time independent case, where the entanglement does not
exhibit dynamics because the temporal dependence of the ground state
of the universe is given by a global phase of the wave function.
We have found  that near the critical coupling, the oscillator's Wigner function in the time-independent case results in a  delocalisation in phase space,  a fact that we associate with the
 ``maximal mixing" scenario reported in  \cite{brandes2} via the divergence of the von Neumann entropy. We have given a detailed prescription for the new entanglement features associated to the system in terms of the linear entropy and its relation to the Wigner function.
The auxiliary dynamical parameters and its stability
properties play a crucial role when determining the entanglement properties of the system and
the behaviour of the universe ground state wave packet.
\begin{acknowledgments}
We acknowledge partial financial support from COLCIENCIAS  under contract
1106-45-221296, and the scientific exchange program PROCOL
(DAAD-Colciencias).
\end{acknowledgments}
\appendix
\section{Solution of the Schr\"odinger equation for a time--depedent quadratic Hamiltonian}
We use the Lewis--Riesenfeld canonical method to find the exact
solution of the Schr\"{o}dinger equation for a system described by
the Hamiltonian Eq. \eqref{GENHAM}. 
With the purpose of building the invariant operator, we define the time--dependent ladder operators
\begin{equation}
\label{ANT}
\hat{a}_{i}(t)=\sum_{k=1}^{2}(A_{ik}(t)\hat{x}_{k}+B_{ik}(t)\hat{p}_{k})\, ,
\end{equation}
and
\begin{equation}
\label{CRT}
\hat{a}^{\dagger}_{i}(t)=\sum_{k=1}^{2}(A^{\ast}_{ik}(t)\hat{x}_{k}+B^{\ast}_{ik}(t)\hat{p}_{k}) \, ,
\end{equation}
which satisfy the standard bosonic commutation relations
\begin{equation}
\label{BCR1}
[\hat{a}_{m}(t),\hat{a}_{n}(t)]=[\hat{a}^{\dagger}_{m}(t),\hat{a}^{\dagger}_{n}(t)]=0\, ,
\end{equation}
and
\begin{equation}
\label{BCR2}
[\hat{a}_{m}(t),\hat{a}^{\dagger}_{n}(t)]=\delta_{mn}.
\end{equation}
The operators $\hat{a}_{i}(t)$ and $\hat{a}^{\dagger}_{i}(t)$ must
satisfy the equation Eq. \eqref{condition}, a
condition which implies that the \textit{auxiliary dynamical parameters} $A_{ik}$ and $B_{ik}$ satisfy
the following differential equations
\begin{eqnarray}
\ddot{B}_{i1}(t) & = &
\frac{\dot{\mu}_{1}(t)}{\mu_{1}(t)}\dot{B}_{i1}-\mu_{1}(t)\nu_{1}(t)B_{i1}(t)\nonumber
\\&&\,-\mu_{1}(t)\gamma(t)B_{i2}(t)  \label{DEDP1} \, ,\\
 \ddot{B}_{i2}(t) & = &
\frac{\dot{\mu}_{2}(t)}{\mu_{2}(t)}\dot{B}_{i2}-\mu_{2}(t)\nu_{2}(t)B_{i2}(t)\nonumber
\\&&\,-\mu_{2}(t)\gamma(t)B_{i1}(t) \, ,
\  \label{DEDP2}
\end{eqnarray}
and
\begin{equation}
\label{CCDP} \dot{B}_{ik}=-\mu_{k}(t)A_{ik}(t) \, .
\end{equation}
As a consequence of these requirements, the auxiliary dynamical
parameters $B_{ik}(t)$ should satisfy
\begin{eqnarray}
\sum_{k}\left(\frac{\dot{B}_{mk}(t)B_{nk}(t)}{\mu_{k}(t)}-\frac{\dot{B}_{nk}(t)B_{mk}(t)}{\mu_{k}(t)}\right)
& = & 0    \label{WC1} \quad , \\
\sum_{k}\left(\frac{\dot{B}^{\ast}_{mk}(t)B^{\ast}_{nk}(t)}{\mu^{\ast}_{k}(t)}-\frac{\dot{B}^{\ast}_{nk}(t)B^{\ast}_{mk}(t)}{\mu^{\ast}_{k}(t)}\right)
& = & 0    \label{WC2}\quad , \\
\sum_{k}\left(\frac{\dot{B}_{mk}(t)B^{\ast}_{nk}(t)}{\mu_{k}(t)}-\frac{B_{mk}(t)\dot{B}^{\ast}_{nk}(t)}{\mu^{\ast}_{k}(t)}\right)&
= & i\delta_{mn}.  \label{WC3}
\end{eqnarray}
Using Eqs. \eqref{DEDP1} and \eqref{DEDP2}, one can easily show
that Eqs. \eqref{WC1}, \eqref{WC2}, and \eqref{WC3} must be time
independent \cite{Hidalgo, Kim}. \\
Interestingly, the auxiliary dynamical parameters $B_{ik}(t)$ are
solutions of the classical equations of motion for a
non-conservative Hamiltonian system \cite{Kim, Arnold}, as is
clearly seen in Eqs. \eqref{DEDP1} and \eqref{DEDP2}. This fact is
related to the Ehrenfest theorem, because for a particle in a
parabolic potential well, the motion of the center of the wave
packet rigorously obeys the laws of classical mechanics
\cite{Tannor}.
The previous procedure allows us to define the Hermitian invariant
operator $ \hat{O}(t)=\hat{n}_{1}(t)+\hat{n}_{2}(t)$, where
$\hat{n}_{k}(t)=\hat{a}_{k}^{\dagger}(t)\hat{a}_{k}(t)$ are time--dependent number operators. We construct the Fock space consisting
of time--dependent number states $|n_{1},n_{2},t\rangle$, where the
vacuum state $|0,0,t\rangle$ is the only one that is annihilated
by $\hat{a}_{1}(t)$ and $\hat{a}_{2}(t)$. The number state is
obtained by applying $\hat{a}_{1}^{\dagger}(t)$ and
$\hat{a}_{2}^{\dagger}(t)$ to the vacuum state
\begin{equation}
\label{fockspace}
|n_{1},n_{2},t\rangle=\frac{(\hat{a}_{1}^{\dagger}(t))^{n_{1}}(\hat{a}_{2}^{\dagger}(t))^{n_{2}}}{\sqrt{(n_{1})!(n_{2})!}}
|0,0,t\rangle .
\end{equation}
\newpage

\end{document}